\documentclass{article}
\usepackage{graphicx}
\usepackage{amsmath}
\usepackage{amsfonts}
\usepackage{amssymb}

\newtheorem{theorem}{Theorem}

\newtheorem{conjecture}{Conjecture}

\newtheorem{definition}{Definition}

\newtheorem{lemma}{Lemma}

\newenvironment{proof}[1][Proof]{\textbf{#1.} }{\ \rule{0.5em}{0.5em}}

\begin{document}

\title{Towards a characterization of exact symplectic Lie algebras $\frak{g}$ in terms of the invariants for the coadjoint representation}

\author{Rutwig Campoamor-Stursberg\\Laboratoire de Math\'ematiques et Applications\\F.S.T., Universit\'e de Haute Alsace\\
F-68093 Mulhouse (France)\\R.Campoamor@uha.fr}

\date{}

\maketitle

\begin{abstract}
We prove that for any known Lie algebra $\frak{g}$ having none invariants for the coadjoint representation, the absence of invariants is equivalent to the existence of a left invariant exact symplectic structure on the corresponding Lie group $G$. We also show that a nontrivial generalized Casimir invariant constitutes an obstruction for the exactness of a symplectic form, and provide solid arguments to conjecture that a Lie algebra is endowed with an exact symplectic form if and only if all invariants for the coadjoint representation are trivial. We moreover develop a practical criterion that allows to deduce the existence of such a symplectic form on a Lie algebra from the shape of the antidiagonal entries of the associated commutator matrix. In an appendix the classification of Lie algebras satisfying $\mathcal{N}(\frak{g})=0$ in low dimensions is given in tabular form, and their exact symplectic structure is given in terms of the Maurer-Cartan equations.
\end{abstract}

\newpage

\section{Introduction}

In the last decades, the role of symplectic geometry and symplectic structures has increased its importance in both mathematics and physics, to constitute nowadays an essential technique in the description and geometrization of natural phenomena. Thus, for example, the magnetic field in magnetostatics determines the symplectic structure of phase space, while the electromagnetic field describes the symplectic character of a relativistic particle. Representation theory, the orbit method and Lie algebra cohomology have also contributed to the analysis of homogeneous symplectic manifolds, deformation and quantization problems or the structural study of dynamical systems \cite{So,Fo,Un,CG}. It is therefore natural to concentrate on  the study of (connected) Lie groups admitting a left invariant symplectic or k\"ahlerian structure, and on the underlying Lie algebras. Many important geometrical results, such as the 1-1 correspondence between equivalence classes of simply connected symplectic homogeneous spaces of a Lie group and the orbit space of equivalence classes of cocycles of the corresponding Lie algebra, from which the nonexistence of (left invariant) symplectic structures on semisimple Lie groups follows \cite{Chu}, have relevant consequences in those physical applications where symplectic structures play an essential role, like in the geometry of the momentum map.\newline On the other hand, symmetry groups have shown that the study of invariants is an indispensable tool for physics (e.g. in the hadron classification \cite{Ge} or the expression of Hamiltonians in terms of Casimir operators). Besides providing an effective method to label irreducible representations of a Lie algebra and to decompose generic representations into irreducible ones, invariants are also essential in the theory of special functions and the symmetry breaking (relating the invariants of the whole group with those of its subgroups). Invariants have been analyzed in detail in low dimensions and for physically relevant groups like the special affine groups $SA(n,\mathbb{R})$, the similitude group or the conformal group of space-time (\cite{Ar,PWZ,Ne} and references therein). However, except for the class of semisimple Lie algebras, no general procedures, algorithms or formulae exist to describe the number of independent invariants and their form in the general case, and the role of some special kinds of invariants (e.g. rational invariants or harmonics) with respect to observable quantities has not been fully explained yet.   

\medskip

In this paper we analyze an interesting geometrical problem concerning the Lie algebras $\frak{g}$ satisfying the equality $\mathcal{N}(\frak{g})=0$, where $\mathcal{N}$ denotes the number of functionally independent invariants for the coadjoint representation. Such an algebra must obviously be even dimensional; it is therefore reasonable to study the existence of symplectic structures on such algebras. We prove that any known Lie algebra satisfying the equality above is indeed endowed with an exact symplectic form, and also that the existence of nontrivial invariants excludes the possibility of constructing an exact symplectic structure on the corresponding Lie group. We therefore conjecture that the characterization of Lie algebras having none invariants for the coadjoint representation is given by the existence of such a symplectic form. This agrees perfectly with the structure of invariants known for large classes of Lie algebras, such as semisimple algebras or algebras having a nilpotent radical, as well as the existence conditions found by Chu in the beginning 1970s.\newline Physically the problem is of interest when studying complete sets of commuting operators in the enveloping Lie algebra of the symmetry algebra $\frak{g}$ of a physical system, where the Casimir operators of the full symmetry group and certain of its subgroup chains provide the quantum numbers in different states. Specifically this is of importance in understanding why certain reductions do not provide such information, since the obtained sets of invariants are not complete. In this sense the existence of an exact symplectic structure can be interpreted as the obstruction to find complete sets of invariants characterizing the system.  Although a complete classification of Lie algebras with  $\mathcal{N}(\frak{g})=0$ is not possible in arbitrary dimension, the characterization can be formulated in full generality in terms of deformations, since Lie algebras carrying an exact symplectic form have been classified up to contraction.\newline As application we give a matrix criterion which shows the equivalence of the existence of an exact symplectic form and the absence of invariants in dependence of the antidiagonal elements of the commutator matrix associated to a Lie algebra, whenever this matrix has a precise shape.  

\medskip 

Unless otherwise stated, any Lie algebra $\frak{g}$ considered in this work is indecomposable and is defined over the field $\mathbb{R}$ of real numbers. We convene that nonwritten brackets are either zero or obtained by antisymmetry. We also use the Einstein summation convention.

\section{Exact symplectic structures}

An even dimensional Lie group $G$ is said to carry a left invariant symplectic structure if it possesses a left invariant closed $2$-form $\omega$ of maximal rank. By abuse of notation such a group is generally called symplectic (whenever there is no ambiguity concerning the group $Sp(n)$) \cite{Chu}. At the Lie algebra level, the existence of the form $\omega$ on $G$ implies that $\omega_{e}$ is a closed skew-symmetric $2$-form on $\frak{g}$ of maximal rank (the converse also holds). Therefore the existence of a left-invariant symplectic structure reduces to the analysis of closed $2$-forms $\omega$ on $\frak{g}$ satisfying
\begin{equation}
\bigwedge^{n}\omega\neq 0,
\end{equation}
where $2n=\dim(\frak{g})$. We call $\frak{g}$ a symplectic Lie algebra. The most elementary Lie algebra having such a structure is the two dimensional affine Lie algebra $\frak{r}_{2}=\left\{X_{1},X_{2}|\quad [X_{1},X_{2}]=X_{2}\right\}$. Defining $\omega\in\frak{r}_{2}^{*}\wedge\frak{r}_{2}^{*}$ by 
\begin{equation}
\omega=\omega_{1}\wedge\omega_{2},
\end{equation}
we get a closed $2$-form of maximal rank. At this point it is important to make a distinction between symplectic structures on Lie algebras. If we consider the preceding algebra in terms of its Maurer-Cartan equations:
\begin{equation}
\left.
\begin{array}{l}
d\omega_{1}=0\\
d\omega_{2}=-\frac{1}{2}\omega_{1}\wedge\omega_{2}
\end{array}
\right\},
\end{equation}
where $\left\{\omega_{1},\omega_{2}\right\}$ is the dual basis to $\left\{X_{1},X_{2}\right\}$, we see that the symplectic form is indeed exact, i.e., there exists a linear form $\alpha\in\frak{r}_{2}^{*}$ such that $\omega=d\alpha$. This fact will have deep consequences for the structure of Lie algebras, which suggests to introduce the following

\begin{definition}
An $2n$-dimensional Lie algebra $\frak{g}$ is called exact symplectic if there exists a form $\omega\in\frak{g}^{*}\wedge\frak{g}^{*}$  such that 
\begin{enumerate}
\item $\omega=d\alpha$ for some linear form $\alpha\in\frak{g}^{*}$,

\item $\bigwedge^{n}\omega\neq 0$.
\end{enumerate}
\end{definition}

That is, a Lie algebra is exact symplectic if it is symplectic and the form is moreover exact (it is important not to confuse this definition of symplectic Lie algebra with the classical one referring to the simple Lie algebras $\frak{sp}(n)$). So for example any four dimensional nilpotent Lie algebra is symplectic, but not exact symplectic. More precisely, we can state:

\begin{lemma}
A nilpotent Lie algebra $\frak{g}$ cannot be exact symplectic.
\end{lemma}

\begin{proof}
Let $\left\{X_{1},...,X_{2n}\right\}$ be a basis of $\frak{g}$ and $\left\{\omega_{1},...,\omega_{2n}\right\}$ its dual basis. The Maurer-Cartan equations of $\frak{g}$ are given by 
\begin{equation}
\begin{array}{l}
d\omega_{k}=-\frac{1}{2}C_{ij}^{k}\omega_{i}\wedge\omega_{j},
\end{array}
\end{equation}
and since the center $Z(\frak{g})$ is nonzero for being nilpotent, a linear form $\omega_{X}$ associated to a central element $X$ does not appear in (4). As a consequence, the image $d\alpha$ of a linear form $\alpha\in\frak{g}^{*}$ by the exterior differential $d$ involving some  summand of the type $\omega_{X}\wedge\omega_{Y}$ ($Y\in\frak{g}$) cannot be closed. 
\end{proof}

Therefore the exactness of the symplectic form excludes nilpotency. In a more general setting, the structural restrictions of symplectic Lie algebras (not necessarily exact) have been analyzed in detail in \cite{Chu}:

\begin{theorem}
Let $\frak{g}$ be a Lie algebra. Then following conditions hold:
\begin{enumerate}

\item if $\frak{g}$ is unimodular (i.e., $trace[ad(X)]=0\quad \forall X\in\frak{g}$) and symplectic, then $\frak{g}$ is solvable.

\item if $\frak{g}$ is semisimple, then it cannot be symplectic.

\item the semidirect sum $\frak{g}\overrightarrow{\oplus}_{R}\frak{n}$ of a semisimple Lie algebra $\frak{g}$ and a nilpotent Lie algebra $\frak{n}$ ($R$ being a representation of $\frak{g}$) cannot be symplectic. 
\end{enumerate}
\end{theorem}

In particular conditions 2. and 3. of the proposition exclude exact symplectic structures. 
From this result, which follows from the analysis undertaken in \cite{Chu}, we deduce that any symplectic Lie algebra in dimension four must be solvable. The converse is easily seen to be false. If we consider the Lie algebra $A_{4,2}^{1}$ given by:
\[
[X_{1},X_{4}]=X_{1},\quad [X_{2},X_{4}]=X_{2},\quad [X_{3},X_{4}]=X_{2}+X_{3},
\]
it can be verified that any closed form $\omega\in(A_{4,2}^{1})^{*}$ satisfies $\omega\wedge\omega=0$. Remarkable is the impossibility of constructing a symplectic structure (in the above sense) on a semisimple Lie algebra. Further it can be shown that if $H^{2}(\frak{g})\neq 0$ and  $\frak{g}$ carries a symplectic structure, then it is necessarily solvable \cite{Ha}. An obvious but important observation is that symplectic structures are preserved by direct sum, i.e., $\frak{g}=\bigoplus_{i}\frak{g}_{i}$ is symplectic if and only if $\frak{g}_{i}$ carries a symplectic structure for all $i$. The same holds for exact symplectic structures. From this behaviour with respect to the direct sums, it follows that for any direct sum of a semisimple Lie algebra $\frak{s}$ with a solvable Lie algebra $\frak{r}$ (that is, a Lie algebra having a trivial Levi decomposition) which is symplectic we have $\frak{s}=0$. For nontrivial Levi decompositions (that is, where the sum is not direct but semidirect) this implication no longer holds, and such algebras will play an important role in what follows. 
In this paper we are interested on exact symplectic Lie algebras in relation with the number $\mathcal{N}(\frak{g})$ of functionally independent invariants for the coadjoint representation $ad^{*}$. We will see that the assumption $d\alpha=\omega$ on the symplectic form $\omega$ is deeply related with the existence of invariants for this representation, and therefore provides a powerful tool to analyze them. To this extent, we discuss briefly the method used to find these invariants, and in particular the Casimir operators of an algebra \cite{PN}. If $\left\{ X_{1},..,X_{n}\right\} $ is a basis of $\frak{g}$ and $\left\{
C_{ij}^{k}\right\} $ the structure constants over this basis, we can
represent $\frak{g}$ in the space $C^{\infty }\left( \frak{g}^{\ast }\right) 
$ by the differential operators 
\begin{equation}
\widehat{X}_{i}=-C_{ij}^{k}x_{k}\frac{\partial }{\partial x_{j}},
\end{equation}
where $\left[ X_{i},X_{j}\right] =C_{ij}^{k}X_{k}$ \ $\left( 1\leq i<j\leq
n\right) $. 
The operators $\widehat{X}_{i}
$ satisfy the brackets $\left[ \widehat{X}_{i},\widehat{X}_{j}\right]
=C_{ij}^{k}\widehat{X}_{k}$ and define a representation of $\frak{g}$ equivalent to $ad$.
A function $F\in C^{\infty }\left( \frak{g}^{\ast }\right) $ is an invariant
if and only if it is a solution of the system: 
\begin{equation}
\left\{\widehat{X}_{i}F=0,\;1\leq i\leq n\right\}.
\end{equation}
Polynomial solutions of the system correspond to classical Casimir invariants. As known, the system $(6)$ can have solutions which are not polynomials, which leads naturally to enlarge the concept of invariant to "generalized Casimir invariants". These solutions also have a fixed value on each irreducible representation of $\frak{g}$. The system may even have no solutions at all, in which case we say that the invariants of the coadjoint representation are trivial. 
In particular, the cardinal $\mathcal{N}$ of a maximal set of functionally independent solutions of the system is given by: 
\begin{equation}
\dim \,\frak{g}-\sup_{x_{1},..,x_{n}} \left\{
rank\left( C_{ij}^{k}x_{k}\right) _{1\leq i<j\leq \dim \frak{g}}\right\},
\end{equation}
where $\left(C_{ij}^{k}x_{k}\right)$ is the matrix which represents the commutator table of $\frak{g}$ over the basis $\left\{X_{1},..,X_{n}\right\}$. \newline As an example, let us consider the following four dimensional solvable Lie algebra $A_{4,12}$:
\begin{equation}
[X_{1},X_{3}]=X_{1},\quad [X_{2},X_{3}]=-X_{2},\quad [X_{1},X_{4}]=-X_{2},\quad [X_{2},X_{4}]=-X_{1}
\end{equation}
This algebra is isomorphic to the Borel subalgebra of $LO(3,1)$, and is of importance in the group classification of nonlinear heat conductivity equations \cite{Bas}. The matrix associated to the commutators of $A_{4,12}$ is:

\begin{equation}
A:=\left( 
\begin{array}{cccc}
0 & 0 & x_{1} & -x_{2}\\ 
0 & 0 & x_{2} & x_{1} \\ 
-x_{1} & -x_{2} & 0 & 0\\ 
x_{2} & -x_{1} & 0 & 0  
\end{array}
\right) 
\end{equation}
and since $\det(A)=(x_{1}^{2}+x_{2}^{2})^{2}$, we obtain $\mathcal{N}(A_{4,12})=0$.\newline 
Invariants of Lie algebras for the coadjoint representation have been determined only in low dimensions, due to the practical impossibility of classifying solvable Lie algebras in dimension $n\geq 7$ (an algorithm to classify them has recently been proposed in the literature, but the complexity of its implementation reduces considerably its applicability \cite{Car}). Solvable Lie algebras have been classified over the reals up to dimension six \cite{Mu1,Mu2,Tu}, and they invariants have been determined \cite{PWZ,Nd}. Lie algebras with nontrivial Levi part have been classified up to dimension nine by Turkowski \cite{Tu2,Tu3}, and further there are partial classifications for higher dimensions, such as for example the rigid Lie algebras \cite{AG}, whose invariants where determined in \cite{Ca,Ca2}. The invariants of Lie algebras having a rank one Levi subalgebra have been studied in some detail in \cite{Ca3}, and it has been shown there that the representations $R$ of the Levi part $\frak{s}$ play an essential role in the analysis of the invariants of semidirect products $\frak{s}\overrightarrow{\oplus}_{R}\frak{r}$.\newline
As known, quantum numbers and observables for particles having a certain symmetry algebra $\frak{g}$ can be obtained from the knowledge of the subalgebras of $\frak{g}$ and their invariants, which can be represented graphically by a tree. There is another type of subalgebras tree which may also be of some interest for physical applications, formed by all (even dimensional) subalgebras for which the equality $\mathcal{N}(\frak{g})=0$ holds. Considering again the previous example, we can easily see that the subalgebra of $A_{4,12}$ generated by $X_{2}$ and $X_{3}$, which is isomorphic to the affine algebra $\frak{r}_{2}$, has also none invariants. Additionally to this fact, these algebras have another important property in common, namely the existence of an exact symplectic form. Writing these algebras in terms of the Maurer-Cartan equations, we obtain:
\begin{equation}
\left.
\begin{array}{l}
d\omega_{1}=-\frac{1}{2}(\omega_{1}\wedge\omega_{3}+\omega_{2}\wedge\omega_{4})\\
d\omega_{2}=-\frac{1}{2}(\omega_{2}\wedge\omega_{3}-\omega_{1}\wedge\omega_{4})\\
d\omega_{3}=d\omega_{4}=0.
\end{array}
\right\},
\end{equation}

\begin{equation}
\left.
\begin{array}{l}
d\omega_{2}=-\frac{1}{2}(\omega_{2}\wedge\omega_{3})\\
d\omega_{3}=0\\
\end{array}
\right\},
\end{equation}

and $d\omega_{1}$ of (10) and $d\omega_{2}$ of (11) are exact symplectic forms on $A_{4,12}$, respectively $\frak{r}_{2}$.\newline 

One may ask whether this constitutes a mere curiosity or if other Lie algebras known to have only trivial invariants will also be endowed with an exact symplectic form. In the Appendix we have tabulated all real Lie algebras of dimension $n\leq 6$ having none invariants. This classification can be deduced partially from the existing literature \cite{PWZ,Nd,Ca}. In preparing Tables 1.-3., some mistakes found in those references have been corrected. We also remark that among these algebras there is only one having a nonzero Levi subalgebra. This algebra, pointed out in \cite{Ca3}, constitutes the proof that the Levi decomposition does not reduce the study of invariants to its semisimple and solvable parts. Tables 1.-3. constitute the proof of the following assertion:

\begin{theorem}
A Lie algebra $\frak{g}$ of dimension $n\leq 6$ satisfies 
\begin{equation}
\mathcal{N}(\frak{g})=0
\end{equation}
if and only if $\frak{g}$ is exact symplectic.
\end{theorem}

Now we can ask if this property is exclusive of low dimensions or if it also holds for higher dimensions. In dimension $8$ we can partially answer, since no classification of solvable algebras in this dimension exists:

\begin{theorem}
Let $\frak{g}$ be a Lie algebra such that
\begin{enumerate}
\item $\frak{g}$ is solvable rigid of dimension $n\leq 8$, or

\item $\frak{g}$ is of dimension $n\leq 8$ and has nontrivial Levi part, or

\item $\frak{g}$ is solvable with abelian nilradical (i.e., maximal nilpotent ideal).
\end{enumerate}
Then $\frak{g}$ satisfies $(12)$ if and only if $\frak{g}$ is exact symplectic.
\end{theorem}

\begin{proof}
The proof of 1. and 2. follows at once from Tables 4 and 5. As to 3., 
solvable Lie algebras with abelian nilradical and satisfying $(12)$ were classified in \cite{Nd}, and they are of the form:
\begin{equation}
\frak{g}\simeq A_{4,12}\oplus(\dim \frak{g}-\dim \frak{n}-2s)\frak{r}_{2}
\end{equation}
where $s$ is the number of distinct complex conjugate roots of $\frak{g}$ and $\frak{n}$ denotes the nilradical (i.e., the maximal nilpotent ideal). From theorem 2 and the fact that exact symplectic structures are preserved by direct sums, the assertion follows.
\end{proof}

This result gives a quite interesting geometrical property which is independent from any classification procedure of Lie algebras. The theorem does moreover hold for any known particular case satisfying (12), and for the families of algebras found by means of sufficiency criteria developed recently \cite{Ca2,Ca3}. Intuitively the property of being exact symplectic seems to be a solid candidate for providing a characterization of algebras having none invariants for the coadjoint representation. Observe moreover that it is not restricted to a particular class of Lie algebras (like 3. in the preceding theorem), but is simultaneously valid for solvable Lie algebras and algebras with nontrivial Levi decomposition (it turns that these are the only types of Lie algebras from which examples satisfying (12) can be extrated). We must however justify at least that the existence of nontrivial invariants excludes the possibility of an exact symplectic structure:

\begin{theorem}
Let $\frak{g}$ be a $2n$-dimensional Lie algebra such that $\mathcal{N}(\frak{g})>0$. Then for any $2$-form $d\omega$ ($\omega\in \frak{g}^{*}$) following equality holds:
\begin{equation}
\bigwedge^{n}d\omega=0.
\end{equation}
\end{theorem} 

\begin{proof}
If $\frak{g}$ were endowed with an exact symplectic structure, then it is known that $\frak{g}$ is a deformation of an algebra which belongs to the following parametrized family $\frak{g}_{\alpha \beta}$ \cite{G1}:
\begin{equation}
\left. 
\begin{array}{l}
\left[ X_{1},X_{2}\right] =X_{1} \\ 
\left[ X_{2r+1},X_{2r+2}\right] =X_{1},\;1\leq r\leq n-1 \\ 
\left[ X_{2},X_{4k-1}\right] =\alpha _{k}X_{4k-1}+\beta _{k}X_{4k+1},\;k\leq
s \\ 
\left[ X_{2},X_{4k}\right] =\left( -1-\alpha _{k}\right) X_{4k}-\beta
_{k}X_{4k+2},\;k\leq s \\ 
\left[ X_{2},X_{4k+1}\right] =-\beta _{k}X_{4k-1}+\alpha
_{k}X_{4k+1},\;k\leq s \\ 
\left[ X_{2},X_{4k+2}\right] =\beta _{k}X_{4k}+\left( -1-\alpha _{k}\right)
X_{4k+2},\;k\leq s \\ 
\left[ X_{2},X_{4s+2k-1}\right] =-\frac{1}{2}X_{4s+2k-1}+\beta
_{k+s-1}X_{4s+2k},\;2\leq k\leq n-2s \\ 
\left[ X_{2},X_{4s+2k}\right] =-\beta _{k+s-1}X_{4k+2s-1}-\frac{1}{2}%
X_{4s+2k},\;2\leq k\leq n-2s
\end{array}
\right\} ,
\end{equation}
where $0\leq s\leq \left[ \frac{n-1}{2}\right] $ and   $\left( \alpha
_{1},..,\alpha _{s},\beta _{1},..,\beta _{n-1-s}\right) \in \Bbb{R}^{n-1}$. It is straightforward to verify that these Lie algebras all satisfy equation $(12)$. Since $\frak{g}$ contracts to some element of the family, by the properties of Lie algebras contractions we deduce that $\mathcal{N}(\frak{g})=0$. Thus, if $I\neq 0$ is an invariant of $\frak{g}$, the algebra cannot contract to an element of $(15)$, and therefore $\frak{g}$ cannot possess an exact symplectic structure.
\end{proof} 

As a consequence of this result, we are naturally led to ennounce the following

\begin{conjecture}
A Lie algebra $\frak{g}$ satisfies $\mathcal{N}(\frak{g})=0$ if and only if it is endowed with an exact symplectic form $\omega$.
\end{conjecture}

Indeed, only one way must be proven, since the other follows from theorem 4. The problem reduces to prove that if $\frak{g}$ is not endowed with an exact symplectic form, then the coadjoint representation $ad^{*}$ necessarily has nontrivial invariants. To prove this implication a general argument not depending on the classification of Lie algebras should be found, and it should be valid without regard on the particular type of Lie algebra. Moreover, from this argument we should be able to deduce a nontrivial solution of the corresponding system  (6). Unfortunately, up to the case of semisimple Lie algebras, where the Killing form corresponds to such an argument, no such property is yet known for the other classes of Lie algebras not being symplectic. However, there are good theoretical reasons for the conjecture to hold, if we take into account simultaneously some facts concerning the invariants of Lie algebras and exact symplectic structures:

\begin{enumerate}

\item By theorem 1, a semisimple Lie algebra $\frak{g}$ cannot be symplectic in the sense of definition 1. On the other hand, it is well known that such algebras have exactly $r$ functionally independent invariants, where $r$ denotes the rank of $\frak{g}$.

\item All known semidirect products $\frak{s}\overrightarrow{\oplus}_{R}\frak{r}$ of a semisimple Lie algebra $\frak{s}$ and a nilpotent algebra $\frak{r}$, described by the representation $R$ of $\frak{s}$, have been shown to posess nontrivial invariants \cite{Ca3}. Again by theorem 1, this type of algebras cannot be symplectic. This applies in particular to semidirect products of the Heisenberg Lie algebra $\frak{h}_{2n+1}$ with a semisimple Lie algebra, which were analyzed in detail in \cite{Ahn}. 
\end{enumerate}

Although these two important observations cannot be taken as conclusive for the correctedness of the assertion, they at least show that the generic results of Lie algebras having a symplectic structure are in accordance with the results obtained in the study of invariants and the theory of homogeneous spaces \cite{Chu,Ha,Nd,Ca,Ca3}. Moreover, if we assume the equivalence of an exact symplectic structure on $\frak{g}$ and the nonexistence of invariants for the coadjoint representation, we have an alternative tool to classify Lie algebras satisfying $(12)$, at least in low dimensions; for practical purposes it is always easier to find closed forms of maximal rank on a Lie algebra than solving the corresponding system $(6)$.

\section{Applications}
It is in general an unsolved problem to construct explicitely symplectic forms on Lie algebras or groups, since they depend on the particular structure of the algebra. However, some results have been obtained independently of classification results \cite{Ca5} for Lie algebras admitting linear forms which decompose in a certain manner. In this section we obtain, by extrapolation of the preceding results, a matricial criterion to ensure that a Lie algebra admits an exact symplectic algebra. More precisely, we show that if on a basis of the Lie algebra $\frak{g}$ the commutator matrix $A( \frak{g})$ has a certain form, then the algebra admits an exact symplectic form. Moreover, the generic shape of the symplectic form can be read off from the matrix. Before ennouncing this criterion, let us analyze an interesting example which serves as motivation: For $n\geq 2$ let us consider the ($2n$)-dimensional Lie algebra $P_{2n}$ defined by the brackets 
\begin{equation}
\lbrack L_{i},L_{j}\rbrack = (j-i)L_{i+j}
\end{equation}
over the basis $\left\{L_{0},..,L_{2n-1}\right\}$. These algebras are easily obtained from the Lie algebra of polynomial vector fields over the one dimensional torus $S^{1}$ (they are finite dimensional quotient algebras of the Virasoro algebra $Vir$). These algebras are graded, and it can easily be shown that they satisfy equation $(12)$. According to our prediction, they should be exact symplectic. Such a form is given by: 
\begin{equation}
\omega=\sum_{k=0}^{2n-1}\omega_{k}\wedge\omega_{2n-1-k},
\end{equation}
where  $\left\{\omega_{0},..,\omega_{2n-1}\right\}$ is dual to the basis $\left\{L_{0},..,L_{2n-1}\right\}$. The form is closed of maximal rank, and since it is a multiple of the form $d\omega_{2n-1}$, it is exact symplectic. The commutator matrix respect to the chosen basis is 
\begin{equation}
\left(
\begin{array}
[c]{ccccccc}%
0 & l_{1}  & 2l_{2} &  ... & \widehat{3}l_{2n-3}  & \widehat{2}l_{2n-2} & \widehat{1}l_{2n-1}\\
-l_{1} & 0 & l_{3} &  ... & \widehat{4}l_{2n-2}  & \widehat{3}l_{2n-1} & 0\\
-2l_{2}  & -l_{3} & 0 &  ... & \widehat{5}l_{2n-1}  & 0 & 0\\
\vdots & \vdots & \vdots &  & \vdots & \vdots & \vdots\\
-\widehat{3}l_{2n-3}  & -\widehat{4}l_{2n-2}  & \widehat{5}l_{2n-1}  &  ... & 0 & 0 & 0\\
-\widehat{2}l_{2n-2} & -\widehat{3}l_{2n-1} & 0 &  ... & 0 & 0 & 0 \\
-\widehat{1}l_{2n-1} & 0 & 0 &  ... & 0  & 0 & 0
\end{array}
\right)  ,
\end{equation}
where we have used the notation $\widehat{k}:=2n-k$. 
Two facts can be deduced by only inspecting this matrix: on one hand, that the determinant is given by the product 
\begin{equation}
\prod_{k=0}^{n-1}\left(  2k+1\right)^{2}l_{2n-1}^{2n}
\end{equation}
of elements corresponding to the "antidiagonal" $(a_{1,2n},a_{2,2n-1},..,a_{2n-1,2},a_{2n,1})$, and on the other that the exact symplectic form corresponds to the dual of the element $L_{2n-1}$. Observe in particular that the remaining constant strcutures do not play an essential role in the determinant, nor in the form $(17)$ chosen. Inspired by this interesting example, we can generalize it to obtain the following result:

\begin{theorem} 
Let $\frak{g}$ be an even dimensional Lie algebra. Suppose that $\frak{g}$ admits a basis $\left\{X_{1},..,X_{2n}\right\}$ such that the commutator matrix $(C_{ij}^{k}x_{k})$ of $\frak{g}$ over it has the form
\begin{equation}
\left(
\begin{array}
[c]{ccccccc}%
a_{11} & a_{12}  & a_{13} &  ... & a_{1,2n-2}  & a_{1,2n-1} & a_{1,2n}\\
-a_{11} & 0 & a_{23} &  ... & a_{2,2n-2}  & a_{2,2n-1} & 0\\
-a_{13}  & -a_{23} & 0 &  ... & a_{3,2n-2}  & 0 & 0\\
\vdots & \vdots & \vdots &  & \vdots & \vdots & \vdots\\
-a_{1,2n-2}  & -a_{2,2n-2}  & -a_{2,2n-1}  &  ... & 0 & 0 & 0\\
-a_{1,2n-1} & -a_{2,2n-1} & 0 &  ... & 0 & 0 & 0 \\
-a_{1,2n-1} & 0 & 0 &  ... & 0  & 0 & 0
\end{array}
\right)  ,
\end{equation}
where $a_{ij}:=C_{ij}^{k}x_{k}$. Then $\frak{g}$ satisfies $\mathcal{N}(\frak{g})=0$ if and only if $\prod_{k=0}^{n}a_{k,2n+1-k}\neq 0$. In particular, $\frak{g}$ is endowed with an exact symplectic structure.
\end{theorem}

\begin{proof}
The proof is straightforward. We only make the following observation concerning the construction of the exact symplectic form. Since the elements of the "antidiagonal" are all nonzero, it follows that $[X_{k},X_{2n+1-k}]=C_{k,2n+1-k}^{t_{k}}X_{t_{k}}\neq 0,\quad 1\leq k\leq n$. Among the images $\left\{X_{t_{k}}\right\}_{1\leq 2n, 1\leq k\leq n}$, choose a minimal set $\left\{X_{t_{k_{1}}},..,X_{t_{k_{r}}}\right\}_{r<2n}$ of independent elements such that, if  $\left\{\omega_{1},..,\omega_{2n}\right\}$ is the dual basis to $\left\{X_{1},..,X_{2n}\right\}$, we have 
\begin{equation}
\omega=\sum_{q=1}^{r}d\omega_{t_{k_{q}}}=\sum_{s=1}^{n}a_{s}\omega_{s}\wedge\omega_{2n+1-s}+ \theta,
\end{equation}
where $a_{s}\neq 0$ for any $s$ and the 2-form $\theta$ does not contain a summand of the type $\omega_{s}\wedge\omega_{2n+1-s}$. The form $\omega$ is obviously closed and exact since it is a linear combination of the Maurer-Cartan equations of $\frak{g}$. Moreover, due to the decomposition $(21)$ the form is also of maximal rank, thus exact symplectic.
\end{proof}

To better visualize the procedure of the proof, we illustrate the construction of the exact symplectic form by an example: let $\frak{L}$ be the eight dimensional Lie algebra defined over the ordered basis $\left\{X_{0},..,X_{7}\right\}$ by the commutator table 

\begin{equation}
A:=\left( 
\begin{array}{cccccccc}
0 & 0 & 0& x_{3}&0  & x_{5}& 0& \lambda x_{7}\\ 
0 & 0 & 0& 0&x_{4}  & x_{5}& \mu x_{6}& 0\\ 
0 & 0 & 0& 0&\alpha x_{4}  & \alpha x_{5}& 0& 0\\ 
-x_{3} & 0 & 0& 0&x_{5}  & 0& 0& 0\\ 
0 & -x_{4} & -\alpha x_{4}& -x_{5}&0  & 0& 0& 0\\ 
-x_{5} &-x_{5} & -\alpha x_{5}& 0&0  & 0& 0& 0\\ 
0 & -\mu x_{6} & 0& 0&0  & 0& 0& 0\\ 
-\lambda x_{7} & 0 & 0& 0&0  & 0& 0& 0 
\end{array}
\right) 
\end{equation}
 
Obviously $\mathcal{N}(\frak{L})=0$ if and only if $\alpha \mu\neq 0$ and $\alpha \lambda\neq 0$. Since $\det(A)=\alpha^{2}\lambda^{2}\mu^{2}x_{6}^{2}x_{7}^{2}x_{5}^{4}$, it suffices to consider the Maurer-Cartan equations $d\omega_{i}$ for $i=5,6,7$. It follows at once that all three must be considered to obtain a form like in $(21)$. Thus take $\omega:=d\omega_{5}+d\omega_{6}+d\omega_{7}$. This form satisfies 
\begin{equation}
\bigwedge^{4}\omega=24\alpha \lambda \mu \omega_{0}\wedge...\wedge\omega_{8},
\end{equation}
which is nonzero if the parameters do not vanish. 

\medskip

This shows that the symplectic form can be read off from the antidiagonal of the commutator matrix. Moreover, algebras satisfing the assumption of the theorem are also in accordance with the conjecture. In particular, both the inspiring example of the factor algebras of the Virasoro algebra as the preceding one suggest that the theorem can be applied to the analysis of exact symplectic structures on graded Lie algebras.

\section{Conclusions}

We have proven that all known classes of Lie algebras having none invariants for the coadjoint representation have a strong geometrical property in common, namely the existence of an exact form of maximal rank. Since we have seen that nontrivial invariants for $ad^{*}$ exclude the possibility of existence of an exact symplectic structure, we have given solid theoretical arguments to conjecture that this is the characterization of Lie algebras having none invariants  . This is in fully agreement with the structural results obtained by various authors on symplectic Lie algebras from the 1970s onwards. In particular this characterization would imply that any Lie algebra $\frak{g}$ satisfying $\mathcal{N}(\frak{g})=0$ can be expressed as a formal deformation of a parametrized family of Lie algebras $\frak{g}_{\alpha \beta}$. This could provide an effective method to classify these algebras up to isomorphism. Since the cohomology of the algebras $\frak{g}_{\alpha \beta}$ is known \cite{AC1}, this classification could be approached, at least in not too high dimensions. The physical interpretation of the conjecture also presents interest. A priori we can interpret the existence of an exact  symplectic structure on a Lie algebra as an obstruction to find a complete set of invariants, therefore to obtain a satisfactory description of the observables and quantum numbers of a physical system \cite{PWZ}. Another problem that arises from this conjecture is the labelling of  the representations of such algebras. It seems reasonable that for these algebras, where the invariants do not provide information, further characteristics are necessary to describe their representations. It has been pointed out that Lie algebras with nontrivial Levi decomposition play an essential role in the analysis of invariants, and that they depend strongly on the pair formed by the representation of the Levi part acting on the radical and the structure of the latter \cite{Ca3}. This leads us naturally to analyze whether the module action on the radical suffices for itself to impede the construction of an exact symplectic structure. A physically relevant example of this interaction is given by the ten dimensional kinematical Lie algebras. While the original classification of these algebras was based on space isotropy and parity and time reversal invariance, they were later generalized asuming only space isotropy \cite{BaN}. The commutator expression of space isotropy is given by:
\begin{equation}
\left[ J_{i},J_{j}\right] =\varepsilon _{ijk}J_{k},\quad \left[ J_{i},K_{j}%
\right] =\varepsilon _{ijk}K_{k},\quad  \left[ J_{i},P_{j}\right] =\varepsilon
_{ijk}P_{k},\quad \left[ J_{i},H\right] =0 
\end{equation}
where $H$ is the time translation, $P_{i}$ the space translations, $J_{i}
$ the rotations and $K_{i}$ the pure Galilean transformations.
Thus, in terms of representation theory, a kinematical Lie algebra $\frak{g}$ possesses a copy of the rotation algebra $\frak{so}(3)$ generated by the $J_{i}$ as subalgebra (this is not necessarily a Levi subalgebra, as shown by the rank two kinematical algebras), and its representation $R=2ad(\frak{so}(3))\oplus D_{0}$, where $D_{0}$ denotes the trivial representation, acts on the remaining elements. Either from the list of kinematical algebras given in \cite{BaN} or by direct computation (the latter has the advantage that all isomorphism classes can be dealt with simultaneously, without concerning about the rank of the Levi part) it can be shown that this representation $R$ excludes the existence of exact symplectic forms, thus that kinematical Lie algebras have all nonzero invariants (for the algebras satisfying parity and time reversal this being obvious, since they are obtained from the de Sitter algebras). Therefore the space isotropy is itself the obstruction to construct such symplectic forms, independently of the additional physical assumptions, or the precise  structure of the algebra. \newline This example also suggests that the effect of a $\frak{so}(3)$-representation $R$ could be of interest in analyzing the solutions of the vacuum Einstein equations associated to the (multidimensional) cosmological models $\frak{so}(3)\overrightarrow{\oplus}_{R}\frak{r}$. In particular one could see whether exact symplectic forms have some effect on the solutions corresponding to the nonstatic case \cite{Sz}.\newline We have also given a practical matrix criterion for Lie algebras $\frak{g}$ to satisfy equation $(12)$. The advantage of this lies in the possibility of obtaining the exact symplectic form on $\frak{g}$ by a mere inspection of the antidiagonal of the matrix. In a sense the form of the matrix suggests the existence of some grading, thus one problem that arises from the criterion is to see if any algebra $\frak{g}$ satisfying it must be indeed graded, and therefore if the property can be extended to succesive subalgebras to construct a subalgebra chain of exact symplectic Lie algebras. If this were so, maybe some additional information on the completeness of the sets of invariants describing a physical system could be gained, or even for symmetry breaking questions if further interactions are present.  

\medskip

Resuming, the obtained results constitute an first step towards a rigorous and complete study of   
 invariants of Lie algebras in terms of representation theory. At least in the case where equation $(12)$ holds, a geometrical meaning has been given to the absence of invariants, and their dependence on representations of semisimple Lie algebras pointed out, showing that there is a sharp division among the (reducible) representations of algebras: those implying (under some additional assumption) the nonexistence of invariants and the remaining. The next natural  step in this frame is to study whether other properties like the existence of a fundamental set of invariants (or even an integrity basis) formed by Casimir operators (or harmonics) are also conditioned by some geometrical property (e.g. nonexact symplectic forms) or by the representation theory of semisimple Lie algebras in the case of nontrivial Levi decompositions (as happens for example for the special affine Lie algebras $\frak{sa}(n,\mathbb{R})$).   

\subsubsection*{Acknowledgment}
During the preparation of this work, the author was supported by a research fellowship of the Fundacion Ram\'on Areces.

\newpage

\section*{Appendix: Exact symplectic Lie algebras in low dimensions}

In this appendix we present, in tabular form, the classification of solvable Lie algebras in
dimension $n\leq 6$, rigid Lie algebras and algebras having nontrivial Levi decomposition in dimensions $n\leq 8$ such that $(12)$ holds, and the equivalence of $(12)$ with the existence of an exact symplectic structure is proved by exhibiting an exact symplectic form. Only the indecomposable algebras are listed. This classification has partially been considered in the recent literature \cite{Nd,Ca,Ca3,Ca5}. The first column of the table contains the label of the algebra and the restrictions on its parameters (if any). The fourth column gives the form $\omega$ defining the exact symplectic structure, in terms of the Maurer-Cartan equations of the algebra, while the intermediate colums list the nonzero brackets. 

\medskip

With respect to the notations used for these Lie algebras, we must make the following observations:

\begin{enumerate}

\item Four dimensional algebras are noted by $A_{4,j}^{ab..}$, where $j$ denotes the number of the isomorphism class, the superscripts (if any) denote the parameters and the basis is labelled as $\left\{X_{1},..,X_{4}\right\}$. The dual basis is noted as $\left\{\omega_{1},..,\omega_{4}\right\}$. The notation is the same as in \cite{PWZ}.

\item Six dimensional solvable Lie algebras with a four dimensional nilradical (Table 1) are denoted by $N_{6,j}^{ab..}$ over the basis $\left\{X_{1},X_{2},N_{1},..,N_{4}\right\}$, according to the notation used in \cite{Tu,Nd}. The corresponding dual bases are noted by $\left\{\omega_{1},\omega_{2},\eta_{1},..,\eta_{4}\right\}$. We also remark that the algebra $N_{6,35}^{ab}$, which was said to satisfy $\mathcal{N}(N_{6,35}^{ab})=0$ in \cite{Nd}, has two invariants if $a=0$.

\item Six dimensional solvable Lie algebras with a five dimensional nilradical (Tables 2. and 3.) are denoted as \cite{Mu1}, where we have corrected minor mistakes found in that classification. We also have added the algebra found in \cite{Tu3}, which did not appear in \cite{Mu1}. Bases are noted by $\left\{X_{1},..,X_{6}\right\}$, and the corresponding dual bases by $\left\{\omega_{1},..,\omega_{6}\right\}$.

\item The notation for the eight dimensional Lie algebras with nonzero Levi part (Table 4) is taken from \cite{Ca3}, and the labels have been adapted from \cite{Tu2}. Bases are denoted by $\left\{X_{1},..,X_{8}\right\}$, with $\left\{\omega_{1},..,\omega_{8}\right\}$ for the dual.

\item The basis of eight dimensional solvable rigid Lie algebras $\frak{g}$ (in this dimension rigidity is equivalent to impose that $H^{2}(\frak{g},\frak{g})=0$) has been taken from \cite{Ca}. Here the elements $Y_{j}$ denote nilpotent elements, while $V_{i}$ has been used for semisimple elements. Dual elements to the $Y_{j}$ have been denoted by $\omega_{Y_{j}}$. We also remark that the algebra labelled with $\frak{r}_{8}^{22}$ in \cite{Ca} was erroneously listed as having none invariants, while it has two independent invariants. For this reason it has not been included in Table 5.

\end{enumerate}
   
\begin{table}
\caption{Lie algebras in dimension $n\leq 6$}
\begin{tabular}
[c]{llll}%
Name & Brackets & & Form\\\hline
$\frak{r}_{2}$  &$[X_{1},X_{2}] =X_{2}$ &  & $d\omega_{2}$\\

$A_{4,7}  $ & $\left[  X_{2},X_{3}\right]
=X_{1}$ & $\left[  X_{1},X_{4}\right]  =2X_{1}$ & $d\omega_{1}$\\
& $\left[  X_{2},X_{4}\right]  =X_{2}$ & $\left[  X_{3},X_{4}\right]
=X_{2}+X_{3}$ & \\

$A_{4,9}^{b}  $ & $\left[  X_{2},X_{3}\right]
=X_{1}$ & $\left[  X_{1},X_{4}\right]  =(1+b)X_{1}$ & $d\omega_{1}$\\
$-1\leq b\leq 1$& $\left[  X_{2},X_{4}\right]  =X_{2}$ & $\left[  X_{3},X_{4}\right]
=bX_{3}$ & \\

$A_{4,11}^{a}  $ & $\left[  X_{2},X_{3}\right]
=X_{1}$ & $\left[  X_{1},X_{4}\right]  =2aX_{1}$ & $d\omega_{1}$\\
$a\geq 0$& $\left[  X_{2},X_{4}\right]  =aX_{2}-X_{3}$ & $\left[  X_{3},X_{4}\right]
=X_{2}+aX_{3}$ & \\

$A_{4,12}  $ & $\left[  X_{1},X_{3}\right]
=X_{1}$ & $\left[  X_{2},X_{3}\right]  =X_{2}$ & $d\omega_{1}$\\
$$& $\left[  X_{1},X_{4}\right]  =-X_{2}$ & $\left[  X_{2},X_{4}\right]
=X_{1}$ & \\

$N_{6,28}$ & $\left[  N_{2},N_{3}\right]
=N_{1}$ & $\left[ N_{3},N_{4}\right]  =N_{2}$ & $d\eta_{1}+d\eta_{3}$\\
& $\left[  X_{1},N_{1}\right]  =N_{1}$ & $\left[X_{1},N_{3}\right]
=-N_{3}$ & \\
& $\left[X_{2},N_{2}\right] =N_{2}$ &
$\left[  X_{2},N_{3}\right]  =2N_{3}$ & \\
& $\left[  X_{1},N_{4}\right]=N_{4}$ & $\left[  X_{2},N_{4}\right]=-N_{4}$ & \\

$N_{6,29}^{a,b}  $ & $\left[  N_{2},N_{3}\right]
=N_{1}$ & $\left[  X_{1},N_{1}\right]  =N_{1}$ & $d\eta_{1}+d\eta_{2}$\\
$a^2+b^2\neq 0$& $\left[  X_{1},N_{2}\right]  =N_{2}$ & $\left[X_{2},N_{3}\right]
=N_{3}$ & \\
& $\left[X_{1},N_{4}\right] =aN_{4}$ &
$\left[  X_{2},N_{1}\right]  =N_{1}$ & \\
& $\left[  X_{2},N_{4}\right]=bN_{4}$ &  & \\

$N_{6,30}^{a}  $ & $\left[  N_{2},N_{3}\right]
=N_{1}$ & $\left[  X_{1},N_{1}\right]  =2N_{1}$ & $d\eta_{1}-d\eta_{3}-d\eta_{4}$\\
$$& $\left[  X_{1},N_{2}\right]  =N_{2}$ & $\left[X_{1},N_{3}\right]
=N_{3}$ & \\
& $\left[X_{1},N_{4}\right] =aN_{4}$ &
$\left[  X_{2},N_{2}\right]  =N_{3}$ & \\
& $\left[  X_{2},N_{4}\right]=N_{4}$ &  & \\

$N_{6,32}^{a}  $ & $\left[  N_{2},N_{3}\right]
=N_{1}$ & $\left[  X_{1},N_{4}\right]  =N_{1}$ & $d\eta_{1}$\\
$$& $\left[  X_{1},N_{2}\right]  =N_{2}$ & $\left[X_{1},N_{3}\right]
=-N_{3}$ & \\
& $\left[X_{2},N_{2}\right] =aN_{2}$ &
$\left[  X_{2},N_{3}\right]  =(1-a)N_{3}$ & \\
& $\left[  X_{2},N_{4}\right]=N_{4}$ &  & \\

$N_{6,33}$ & $\left[  N_{2},N_{3}\right]
=N_{1}$ & $\left[  X_{1},N_{2}\right]  =N_{2}$ & $d\eta_{1}+d\eta_{4}$\\
$$& $\left[  X_{1},N_{1}\right]  =N_{1}$ & $\left[X_{2},N_{4}\right]
=N_{4}$ & \\
& $\left[X_{2},N_{3}\right] =N_{3}+N_{4}$ &
$\left[  X_{2},N_{1}\right]  =N_{1}$ & \\

$N_{6,34}^{a}  $ & $\left[  N_{2},N_{3}\right]
=N_{1}$ & $\left[  X_{1},N_{1}\right]  =N_{1}$ & $d\eta_{1}+d\eta_{4}$\\
$$& $\left[  X_{1},N_{3}\right]  =N_{4}$ & $\left[X_{2},N_{1}\right]
=(1+a)N_{1}$ & \\
& $\left[X_{2},N_{2}\right] =aN_{2}$ &
$\left[  X_{2},N_{3}\right]  =N_{3}$ & \\
& $\left[  X_{1},N_{2}\right]=N_{2}$ & $\left[X_{2},N_{4}\right]=N_{4}$ & \\
$N_{6,35}^{a,b}  $ & $\left[  N_{2},N_{3}\right]
=N_{1}$ & $\left[  X_{1},N_{2}\right]  =N_{3}$ & $d\eta_{1}+d\eta_{4}$\\
$a\neq 0, a+b\neq 0$& $\left[  X_{1},N_{4}\right]  =aN_{4}$ & $\left[X_{1},N_{3}\right]
=-N_{2}$ & \\
& $\left[X_{2},N_{1}\right] =2N_{1}$ &
$\left[  X_{2},N_{2}\right]  =N_{2}$ & \\
& $\left[  X_{2},N_{3}\right]=N_{3}$ & $\left[  X_{2},N_{4}\right]=bN_{4}$ & \\

$N_{6,37}^{a}  $ & $\left[  N_{2},N_{3}\right]
=N_{1}$ & $\left[  X_{2},N_{1}\right]  =2N_{1}$ & $d\eta_{1}$\\
$$& $\left[  X_{1},N_{3}\right]  =-N_{2}$ & $\left[X_{2},N_{4}\right]
=2N_{4}$ & \\
& $\left[X_{1},N_{2}\right] =N_{3}$ &
$\left[  X_{2},N_{3}\right]  =-aN_{2}+N_{3}$ & \\
& $\left[  X_{2},N_{2}\right]=N_{2}+aN_{3}$ & $\left[X_{1},N_{4}\right]=N_{1}$ & \\\hline
\end{tabular}
\end{table}

\begin{table}

\caption{Lie algebras in dimension $n\leq 6$}
\begin{tabular}
[c]{llll}%
Name & Brackets & & Form\\\hline
$\frak{g}_{6,82}  $ & $\left[  X_{2},X_{4}\right]
=X_{1}$ & $\left[  X_{3},X_{5}\right]  =X_{1}$ & $d\omega_{1}$\\
$\alpha=2$& $\left[  X_{1},X_{6}\right]  =2X_{1}$ & $\left[  X_{2},X_{6}\right]
=\left(  \lambda+1\right)  X_{2}$ & \\
& $\left[  X_{3},X_{6}\right]  =\left(  \lambda_{1}+1\right)  X_{3}$ &
$\left[  X_{4},X_{6}\right]  =\left(  -\lambda+1\right)  X_{4}$ & \\
& $\left[  X_{5},X_{6}\right]  =\left(  -\lambda_{1}+1\right)  X_{5}$ &  & \\
$\frak{g}_{6,83}$ & $\left[  X_{2},X_{4}\right]
=X_{1}$ & $\left[  X_{3},X_{5}\right]  =X_{1}$ & $d\omega_{1}$\\
$\alpha\neq0$& $\left[  X_{1},X_{6}\right]  =\alpha X_{1}$ & $\left[  X_{2},X_{6}\right]
=\left(  \frac{\alpha}{2}+\lambda\right)  X_{2}$ & \\
& $\left[  X_{3},X_{6}\right]  =\left(  \frac{\alpha}{2}+\lambda\right)
X_{3}$ & $\left[  X_{4},X_{6}\right]  =\left(  \frac{\alpha}{2}-\lambda
\right)  X_{4}$ & \\
& $\left[  X_{5},X_{6}\right]  =-X_{4}+\left(  \frac{\alpha}{2}-\lambda
\right)  X_{5}$ &  & \\
$\frak{g}_{6,85}$ & $\left[  X_{2},X_{4}\right]  =X_{1}$ & $\left[
X_{3},X_{5}\right]  =X_{1}$ & $d\omega_{1}$\\
& $\left[  X_{1},X_{6}\right]  =2X_{1}$ & $\left[  X_{2},X_{6}\right]
=\left(  1+\lambda\right)  X_{2}$ & \\
& $\left[  X_{3},X_{6}\right]  =X_{3}$ & $\left[  X_{4},X_{6}\right]  =\left(
1-\lambda\right)  X_{4}$ & \\
& $\left[  X_{5},X_{6}\right]  =X_{3}+X_{5}$ &  & \\
$\frak{g}_{6,86}$ & $\left[  X_{2},X_{4}\right]  =X_{1}$ & $\left[
X_{3},X_{5}\right]  =X_{1}$ & $d\omega_{1}$\\
& $\left[  X_{1},X_{6}\right]  =2X_{1}$ & $\left[  X_{2},X_{6}\right]
=X_{2}+X_{3}$ & \\
& $\left[  X_{3},X_{6}\right]  =X_{3}$ & $\left[  X_{4},X_{6}\right]  =X_{4}$
& \\
& $\left[  X_{5},X_{6}\right]  =-X_{4}+X_{5}$ &  & \\
$\frak{g}_{6,87}$ & $\left[  X_{2},X_{4}\right]  =X_{1}$ & $\left[
X_{3},X_{5}\right]  =X_{1}$ & $d\omega_{1}$\\
& $\left[  X_{1},X_{6}\right]  =2X_{1}$ & $\left[  X_{2},X_{6}\right]
=X_{2}+X_{5}$ & \\
& $\left[  X_{3},X_{6}\right]  =X_{3}+X_{4}$ & $\left[  X_{4},X_{6}\right]
=X_{4}$ & \\
& $\left[  X_{5},X_{6}\right]  =X_{3}+X_{5}$ &  & \\
$\frak{g}_{6,88}$ & $\left[  X_{2},X_{4}\right]
=X_{1}$ & $\left[  X_{3},X_{5}\right]  =X_{1}$ & $d\omega_{1}$\\
$\alpha\neq0$& $\left[  X_{1},X_{6}\right]  =\alpha X_{1}$ & $\left[  X_{2},X_{6}\right]
=\left(  \frac{\alpha}{2}+\mu_{0}\right)  X_{2}+\upsilon_{0}X_{3}$ & \\
& $\left[  X_{3},X_{6}\right]  =-\upsilon_{0}X_{2}+\left(  \frac{\alpha}
{2}+\mu_{0}\right)  X_{3}$ & $\left[  X_{4},X_{6}\right]  =\left(
\frac{\alpha}{2}-\mu_{0}\right)  X_{4}+\upsilon_{0}X_{5}$ & \\
& $\left[  X_{5},X_{6}\right]  =-\upsilon_{0}X_{4}+\left(  \frac{\alpha}
{2}-\mu_{0}\right)  X_{5}$ &  & \\
$\frak{g}_{6,89} $ & $\left[  X_{2},X_{4}\right]
=X_{1}$ & $\left[  X_{3},X_{5}\right]  =X_{1}$ & $d\omega_{1}$\\
$\alpha\neq0$& $\left[  X_{1},X_{6}\right]  =\alpha X_{1}$ & $\left[  X_{2},X_{6}\right]
=\left(  \frac{\alpha}{2}+s\right)  X_{2}$ & \\
& $\left[  X_{3},X_{6}\right]  =\upsilon_{0}X_{5}+\frac{\alpha}{2}X_{3}$ &
$\left[  X_{4},X_{6}\right]  =\left(  \frac{\alpha}{2}-s\right)  X_{4}$ & \\
& $\left[  X_{5},X_{6}\right]  =-\upsilon_{0}X_{3}+\frac{\alpha}{2}X_{5}$ &
& \\
$\frak{g}_{6,90}$ & $\left[  X_{2},X_{4}\right]
=X_{1}$ & $\left[  X_{3},X_{5}\right]  =X_{1}$ & $d\omega_{1}$\\
$\alpha\neq0$& $\left[  X_{1},X_{6}\right]  =\alpha X_{1}$ & $\left[  X_{2},X_{6}\right]
=\frac{\alpha}{2}X_{2}+X_{4}$ & \\
& $\left[  X_{3},X_{6}\right]  =\frac{\alpha}{2}X_{3}+v_{0}X_{5}$ & $\left[
X_{4},X_{6}\right]  =X_{2}+\frac{\alpha}{2}X_{4}$ & \\
& $\left[  X_{5},X_{6}\right]  =-\upsilon_{0}X_{3}+\frac{\alpha}{2}X_{5}$ &
& \\
$\frak{g}_{6,92}$ & $\left[  X_{2},X_{4}\right]
=X_{1}$ & $\left[  X_{3},X_{5}\right]  =X_{1}$ & $d\omega_{1}$\\
$\alpha\neq0$& $\left[  X_{1},X_{6}\right]  =\alpha X_{1}$ & $\left[  X_{2},X_{6}\right]
=\frac{\alpha}{2}X_{2}+\upsilon_{0}X_{3}$ & \\
& $\left[  X_{3},X_{6}\right]  =-\mu_{0}X_{2}+\frac{\alpha}{2}X_{3}$ &
$\left[  X_{4},X_{6}\right]  =\mu_{0}X_{5}+\frac{\alpha}{2}X_{4}$ & \\
& $\left[  X_{5},X_{6}\right]  =-\upsilon_{0}X_{4}+\frac{\alpha}{2}X_{5}$ &
& \\
$\frak{g}_{6,92}^{*}$ & $\left[  X_{2},X_{4}\right]
=X_{5}$ & $\left[  X_{1},X_{3}\right]  =X_{5}$ & $d\omega_{5}$\\
$p\neq0$& $\left[  X_{1},X_{6}\right]  =pX_{1}+X_{3}$ & $\left[  X_{2},X_{6}\right]
=pX_{2}+X_{4}$ & \\
& $\left[  X_{3},X_{6}\right]  =-X_{1}+pX_{3}$ &
$\left[  X_{4},X_{6}\right]  =-X_{2}+pX_{4}$ & \\
& $\left[  X_{5},X_{6}\right]  =2pX_{5}$ &
& \\\hline
\end{tabular}
\end{table}

\begin{table}
\caption{Lie algebras in dimension $n\leq 6$}
\begin{tabular}
[c]{llll}%
Name & Brackets & & Form\\\hline
$\frak{g}_{6,93} $ & $\left[  X_{2},X_{4}\right]
=X_{1}$ & $\left[  X_{3},X_{5}\right]  =X_{1}$ & $d\omega_{1}$\\
$\alpha\neq0$& $\left[  X_{1},X_{6}\right]  =\alpha X_{1}$ & $\left[  X_{2},X_{6}\right]
=\frac{\alpha}{2}X_{2}+X_{4}+\upsilon_{0}X_{5}$ & \\
& $\left[  X_{3},X_{6}\right]  =\upsilon_{0}X_{4}+\frac{\alpha}{2}X_{3}$ &
$\left[  X_{4},X_{6}\right]  =X_{2}-\upsilon_{0}X_{3}+\frac{\alpha}{2}X_{4}$ &
\\
& $\left[  X_{5},X_{6}\right]  =-\upsilon_{0}X_{2}+\frac{\alpha}{2}X_{5}$ &
& \\
$\frak{g}_{6,94} $ & $\left[  X_{3}%
,X_{4}\right]  =X_{1}$ & $\left[  X_{2},X_{5}\right]  =X_{1}$ & $d\omega_{1}$\\
$\lambda+2\neq0$& $\left[  X_{3},X_{5}\right]  =X_{2}$ & $\left[  X_{1},X_{6}\right]  =\left(
\lambda+2\right)  X_{1}$ & \\
& $\left[  X_{2},X_{6}\right]  =\left(  \lambda+1\right)  X_{2}$ & $\left[
X_{3},X_{6}\right]  =\lambda X_{3}$ & \\
& $\left[  X_{4},X_{6}\right]  =2X_{4}$ & $\left[  X_{5},X_{6}\right]  =X_{5}$
& \\
$\frak{g}_{6,95}$ & $\left[  X_{3},X_{4}\right]  =X_{1}$ & $\left[
X_{2},X_{5}\right]  =X_{1}$ & $d\omega_{1}$\\
& $\left[  X_{3},X_{5}\right]  =X_{2}$ & $\left[  X_{1},X_{6}\right]  =2X_{1}$
& \\
& $\left[  X_{2},X_{6}\right]  =X_{2}$ & $\left[  X_{3},X_{6}\right]  =0$ & \\
& $\left[  X_{4},X_{6}\right]  =X_{1}+2X_{4}$ & $\left[  X_{5},X_{6}\right]
=X_{5}$ & \\
$\frak{g}_{6,96}$ & $\left[  X_{3},X_{4}\right]  =X_{1}$ & $\left[
X_{2},X_{5}\right]  =X_{1}$ & $d\omega_{1}$\\
& $\left[  X_{3},X_{5}\right]  =X_{2}$ & $\left[  X_{1},X_{6}\right]  =3X_{1}$
& \\
& $\left[  X_{2},X_{6}\right]  =2X_{2}$ & $\left[  X_{3},X_{6}\right]  =X_{3}$
& \\
& $\left[  X_{4},X_{6}\right]  =X_{2}+2X_{4}$ & $\left[  X_{5},X_{6}\right]
=X_{3}+X_{5}$ & \\
$\frak{g}_{6,97}$ & $\left[  X_{3},X_{4}\right]  =X_{1}$ & $\left[
X_{2},X_{5}\right]  =X_{1}$ & $d\omega_{1}$\\
& $\left[  X_{3},X_{5}\right]  =X_{2}$ & $\left[  X_{1},X_{6}\right]  =4X_{1}$
& \\
& $\left[  X_{2},X_{6}\right]  =3X_{2}$ & $\left[  X_{3},X_{6}\right]
=2X_{3}+X_{4}$ & \\
& $\left[  X_{4},X_{6}\right]  =2X_{4}$ & $\left[  X_{5},X_{6}\right]  =X_{5}$
& \\
$\frak{g}_{6,98}$ & $\left[  X_{3},X_{4}\right]  =X_{1}$ & $\left[
X_{2},X_{5}\right]  =X_{1}$ & $d\omega_{1}$\\
& $\left[  X_{3},X_{5}\right]  =X_{2}$ & $\left[  X_{1},X_{6}\right]  =X_{1}$
& \\
& $\left[  X_{2},X_{6}\right]  =hX_{1}+X_{2}$ & $\left[  X_{3},X_{6}\right]
=X_{3}$ & \\
& $\left[  X_{4},X_{6}\right]  =0$ & $\left[  X_{5},X_{6}\right]  =hX_{4}$ & \\
$\frak{g}_{6,99}$ & $\left[  X_{3},X_{4}\right]  =X_{1}$ & $\left[
X_{2},X_{5}\right]  =X_{1}$ & $d\omega_{1}$\\
& $\left[  X_{3},X_{5}\right]  =X_{2}$ & $\left[  X_{4},X_{5}\right]  =X_{3}$
& \\
& $\left[  X_{1},X_{6}\right]  =5X_{1}$ & $\left[  X_{2},X_{6}\right]
=4X_{2}$ & \\
& $\left[  X_{3},X_{6}\right]  =3X_{3}$ & $\left[  X_{4},X_{6}\right]
=2X_{4}$ & \\
& $\left[  X_{5},X_{6}\right]  =X_{5}$ &  & \\
$L_{6}$ & $\left[  X_{1},X_{2}\right]  =2X_{2}$ & $\left[
X_{1},X_{3}\right]  =-2X_{3}$ & $d\omega_{1}+d\omega_{4}+d\omega_{5}$\\
& $\left[  X_{2},X_{3}\right]  =X_{1}$ & $\left[  X_{1},X_{4}\right]  =X_{4}$
& \\
& $\left[  X_{1},X_{5}\right]  =X_{5}$ & $\left[  X_{2},X_{5}\right]
=X_{4}$ & \\
& $\left[  X_{3},X_{4}\right]  =X_{5}$ & $\left[  X_{4},X_{6}\right]
=X_{4}$ & \\
& $\left[  X_{5},X_{6}\right]  =X_{5}$ &  & \\\hline

\end{tabular}
\end{table}

\begin{table}
\caption{Lie algebras with nonzero Levi part (dimension 8)}
\begin{tabular}
[c]{lllll}%
Name & Brackets &  & & Form\\\hline
$L_{8,3} $ & $\left[ X_{2},X_{3}\right]
=X_{1}$ & $\left[  X_{1},X_{2}\right]  =X_{3}$ & $\left[  X_{1},X_{3}\right]  =-X_{2}$ & $d\omega_{4}$\\
 & $\left[  X_{1},X_{4}\right]
=\frac{1}{2}X_{7}$ & $\left[  X_{1},X_{5}\right]  =\frac{1}{2}X_{6}$ &
$\left[  X_{1},X_{6}\right]  =-\frac{1}{2}X_{5}$ & \\
& $\left[X_{1},X_{7}\right]=-\frac{1}{2}X_{4}$ & $\left[X_{2},X_{4}\right]=\frac{1}{2}X_{5}$& $\left[X_{2},X_{5}\right]=-\frac{1}{2}X_{4}$&\\
& $\left[X_{2},X_{6}\right]=\frac{1}{2}X_{7}$ & $\left[X_{2},X_{7}\right]=-\frac{1}{2}X_{6}$
& $\left[X_{3},X_{4}\right]=\frac{1}{2}X_{6}$ & \\
& $\left[X_{3},X_{5}\right]=-\frac{1}{2}X_{7}$ & $\left[X_{3},X_{6}\right]=-\frac{1}{2}X_{4}$ & $\left[X_{3},X_{7}\right]=\frac{1}{2}X_{5}$ & \\
& $\left[X_{4},X_{8}\right]=X_{4}$ & $\left[X_{5},X_{8}\right]=X_{5}$ & $\left[X_{6},X_{8}\right]=X_{6}$ & \\
& $\left[X_{7},X_{8}\right]=X_{7}$ & & &\\

$L_{8,4}^{p} $ & $\left[ X_{2},X_{3}\right]
=X_{1}$ & $\left[  X_{1},X_{2}\right]  =X_{3}$ & $\left[  X_{1},X_{3}\right]  =-X_{2}$ & $d\omega_{7}$\\
 & $\left[  X_{1},X_{4}\right]
=\frac{1}{2}X_{7}$ & $\left[  X_{1},X_{5}\right]  =\frac{1}{2}X_{6}$ &
$\left[  X_{1},X_{6}\right]  =-\frac{1}{2}X_{5}$ & \\
& $\left[X_{1},X_{7}\right]=-\frac{1}{2}X_{4}$ & $\left[X_{2},X_{4}\right]=\frac{1}{2}X_{5}$& $\left[X_{2},X_{5}\right]=-\frac{1}{2}X_{4}$&\\
& $\left[X_{2},X_{6}\right]=\frac{1}{2}X_{7}$ & $\left[X_{2},X_{7}\right]=-\frac{1}{2}X_{6}$
& $\left[X_{3},X_{4}\right]=\frac{1}{2}X_{6}$ & \\
& $\left[X_{3},X_{5}\right]=-\frac{1}{2}X_{7}$ & $\left[X_{3},X_{6}\right]=-\frac{1}{2}X_{4}$ & $\left[X_{3},X_{7}\right]=\frac{1}{2}X_{5}$ & \\
& $\left[X_{4},X_{8}\right]=pX_{4}-X_{6}$ & $\left[X_{5},X_{8}\right]=pX_{5}-X_{7}$ & $\left[X_{6},X_{8}\right]=X_{4}+pX_{6}$ & \\
& $\left[X_{7},X_{8}\right]=X_{5}+pX_{7}$ & & &\\

$L_{8,16} $ & $\left[ X_{2},X_{3}\right]
=X_{1}$ & $\left[  X_{1},X_{2}\right]  =2X_{2}$ & $\left[  X_{1},X_{3}\right]  =-2X_{3}$ & $d\omega_{5}+d\omega_{7}$\\
 & $\left[  X_{1},X_{4}\right]
=X_{4}$ & $\left[  X_{1},X_{5}\right]  =-X_{5}$ &
$\left[  X_{1},X_{6}\right]  =X_{6}$ & \\
& $\left[X_{1},X_{7}\right]=-X_{7}$ & $\left[X_{2},X_{5}\right]=X_{4}$& $\left[X_{2},X_{7}\right]=X_{6}$&\\
& $\left[X_{3},X_{4}\right]=X_{5}$ & $\left[X_{3},X_{6}\right]=X_{7}$
& $\left[X_{4},X_{8}\right]=X_{4}$ & \\
& $\left[X_{5},X_{8}\right]=X_{5}$ & $\left[X_{6},X_{8}\right]=X_{4}+X_{6}$ & $\left[X_{7},X_{8}\right]=X_{5}+X_{7}$ & \\

$L_{8,17}^{p} $ & $\left[ X_{2},X_{3}\right]
=X_{1}$ & $\left[  X_{1},X_{2}\right]  =2X_{2}$ & $\left[  X_{1},X_{3}\right]  =-2X_{3}$ & $d\omega_{4}+d\omega_{5}$\\
$p\neq -1$ & $\left[  X_{1},X_{4}\right]
=X_{4}$ & $\left[  X_{1},X_{5}\right]  =-X_{5}$ &
$\left[  X_{1},X_{6}\right]  =X_{6}$ & \\
& $\left[X_{1},X_{7}\right]=-X_{7}$ & $\left[X_{2},X_{5}\right]=X_{4}$& $\left[X_{2},X_{7}\right]=X_{6}$&\\
& $\left[X_{3},X_{4}\right]=X_{5}$ & $\left[X_{3},X_{6}\right]=X_{7}$
& $\left[X_{4},X_{8}\right]=X_{4}$ & \\
& $\left[X_{5},X_{8}\right]=X_{5}$ & $\left[X_{6},X_{8}\right]=pX_{6}$ & $\left[X_{7},X_{8}\right]=pX_{7}$ & \\

$L_{8,18}^{p} $ & $\left[ X_{2},X_{3}\right]
=X_{1}$ & $\left[  X_{1},X_{2}\right]  =2X_{2}$ & $\left[  X_{1},X_{3}\right]  =-2X_{3}$ & $d\omega_{4}+d\omega_{7}$\\
$p\neq 0$ & $\left[  X_{1},X_{4}\right]
=X_{4}$ & $\left[  X_{1},X_{5}\right]  =-X_{5}$ &
$\left[  X_{1},X_{6}\right]  =X_{6}$ & \\
& $\left[X_{1},X_{7}\right]=-X_{7}$ & $\left[X_{2},X_{5}\right]=X_{4}$& $\left[X_{2},X_{7}\right]=X_{6}$&\\
& $\left[X_{3},X_{4}\right]=X_{5}$ & $\left[X_{3},X_{6}\right]=X_{7}$
& $\left[X_{4},X_{8}\right]=pX_{4}-X_{6}$ & \\
& $\left[X_{5},X_{8}\right]=pX_{5}-X_{7}$ & $\left[X_{6},X_{8}\right]=X_{4}+pX_{6}$ & $\left[X_{7},X_{8}\right]=X_{5}+pX_{7}$ & \\

$L_{8,20} $ & $\left[ X_{2},X_{3}\right]
=X_{1}$ & $\left[  X_{1},X_{2}\right]  =2X_{2}$ & $\left[  X_{1},X_{3}\right]  =-2X_{3}$ & $d\omega_{4}-d\omega_{7}$\\
 & $\left[  X_{1},X_{4}\right]
=3X_{4}$ & $\left[  X_{1},X_{5}\right]  =X_{5}$ &
$\left[  X_{1},X_{6}\right]  =-X_{6}$ & \\
& $\left[X_{1},X_{7}\right]=-3X_{7}$ & $\left[X_{2},X_{5}\right]=3X_{4}$& $\left[X_{2},X_{6}\right]=2X_{5}$&\\
& $\left[X_{2},X_{7}\right]=X_{6}$ & $\left[X_{3},X_{4}\right]=X_{4}$
& $\left[X_{3},X_{5}\right]=2X_{6}$ & \\
& $\left[X_{3},X_{6}\right]=3X_{7}$ & $\left[X_{4},X_{8}\right]=X_{4}$ & $\left[X_{5},X_{8}\right]=X_{5}$ & \\
& $\left[X_{6},X_{8}\right]=X_{6}$ & $\left[X_{7},X_{8}\right]=X_{7}$ &  & \\\hline
\end{tabular}
\end{table}

\begin{table}[h]
\caption{Solvable rigid Lie algebras in dimension $8$ }
\begin{tabular}
[c]{llll}%
Name & Brackets & & Form\\\hline
$\frak{g}_{8}^{1}$ & $\left[  V_{1},Y_{i}\right]

=iY_{i}, i=1,2,3,5,6,7,8$ & $\left[Y_{1},Y_{i}\right] =Y_{i+1}, i=2,6,7$ & $d\omega_{y_{8}}$\\

& $\left[Y_{2},Y_{3}\right] =Y_{5}$& $\left[  Y_{2},Y_{5}\right]  =Y_{7}$ & \\
& $\left[ Y_{2},Y_{6}\right]=Y_{8}$ & $\left[  Y_{3},Y_{5}\right]  =Y_{7}$ & \\

$\frak{g}_{8}^{6}$ & $\left[  V_{1},Y_{i}\right]

=iY_{i}, i=2,3,4,6,7,8,10$ & $\left[Y_{2},Y_{i}\right] =Y_{i+2}, i=4,6,8$ & $d\omega_{y_{10}}$\\

& $\left[Y_{3},Y_{i}\right] =Y_{i+3}? i=4,7$& $\left[  Y_{4},Y_{6}\right]  =Y_{10}$ & \\

$\frak{g}_{8}^{9}$ & $\left[  V_{1},Y_{i}\right]

=iY_{i}, i=1,2,3,4,5,6$ & $\left[V_{1},Y^{\prime}_{3}\right] =3Y^{\prime}_{3}$ & $d\omega_{y_{6}}$\\

& $\left[Y_{1},Y_{i+1}\right] =Y_{i+1}? i=2,3,4,5$& $\left[  Y_{2},Y_{i}\right]  =Y_{i+2}; i=3,4$ & \\
& $\left[ Y_{2},Y^{\prime}_{3}\right]=Y_{5}$ & $\left[  Y_{3},Y^{\prime}_{3}\right]  =Y_{6}$ & \\

$\frak{g}_{8}^{20}$ & $\left[  V_{1},Y_{i}\right]

=iY_{i}, i=1,2,3,4,5,6$ & $\left[V_{2},Y_{i}\right] =Y_{i}, i=3,4,5,6$ & $d\omega_{y_{3}}+d\omega_{y_{6}}$\\

& $\left[Y_{1},Y_{i}\right] =Y_{i+1}, i=3,4,5$& $\left[  Y_{2},Y_{i}\right]  =Y_{i+2}, i=3,4$ & \\

$\frak{g}_{8}^{21}$ & $\left[  V_{1},Y_{i}\right]

=iY_{i}, i=1,2,3,4,5,6$ & $\left[V_{2},Y_{i}\right] =Y_{i}, i=4,5,6$ & $d\omega_{y_{3}}+d\omega_{y_{6}}$\\

& $\left[Y_{1},Y_{i}\right] =Y_{i+1}, i=2,4,5$& $\left[  Y_{2},Y_{4}\right]  =Y_{6}$ & \\

$\frak{g}_{8}^{23}$ & $\left[  V_{1},Y_{i}\right]

=iY_{i}, i=1,2,3,4,5,6$ & $\left[V_{2},Y_{6}\right] =Y_{6}$ & $d\omega_{y_{5}}+d\omega_{y_{6}}$\\

& $\left[Y_{1},Y_{i}\right] =Y_{i+1}, i=2,3,4$& $\left[  Y_{2},Y_{3}\right]  =Y_{5}$ & \\

$\frak{g}_{8}^{24}$ & $\left[  V_{1},Y_{i}\right]

=iY_{i}, i=1,2,3,4,5,7$ & $\left[V_{2},Y_{i}\right] =Y_{i}, i=3,4,5$ & $d\omega_{y_{5}}+d\omega_{y_{7}}$\\

& $\left[Y_{1},Y_{i}\right] =Y_{i+1}, i=3,4$& $\left[  V_{2},Y_{7}\right]  =2Y_{7}$ & \\

& $\left[Y_{2},Y_{3}\right] =Y_{5}$& $\left[  Y_{3},Y_{4}\right]  =Y_{7}$ & \\

$\frak{g}_{8}^{25}$ & $\left[  V_{1},Y_{i}\right]

=iY_{i}, i=1,2,3,4,5,7$ & $\left[V_{2},Y_{i}\right] =Y_{i}, i=2,3,4$ & $d\omega_{y_{4}}+d\omega_{y_{7}}$\\

& $\left[Y_{1},Y_{i}\right] =Y_{i+1}, i=2,3$& $\left[  V_{2},Y_{5}\right]  =2Y_{5}$ & \\

& $\left[V_{2},Y_{7}\right] =3Y_{7}$& $\left[  Y_{2},Y_{i}\right]  =Y_{i+2}, i=3,5$ & \\

$\frak{g}_{8}^{26}$ & $\left[  V_{1},Y_{i}\right]

=iY_{i}, i=1,2,3,4,5,7$ & $\left[V_{2},Y_{i}\right] =Y_{i}, i=2,3,4,5$ & $d\omega_{y_{1}}+d\omega_{y_{7}}$\\

& $\left[Y_{1},Y_{i}\right] =Y_{i+1}, i=2,3,4$& $\left[  Y_{2},Y_{5}\right]  =Y_{7}$ & \\

& $\left[Y_{3},Y_{4}\right] =-Y_{7}$&  & \\

$\frak{g}_{8}^{27}$ & $\left[  V_{1},Y_{i}\right]

=iY_{i}, i=1,2,3,5,6,7$ & $\left[V_{2},Y_{i}\right] =Y_{i}, i=2,3$ & $d\omega_{y_{3}}+d\omega_{y_{7}}$\\

& $\left[Y_{1},Y_{i}\right] =Y_{i+1}, i=2,6$& $\left[  V_{2},Y_{5}\right]  =2Y_{5}$ & \\

& $\left[V_{2},Y_{6}\right] =3Y_{6}$& $\left[  V_{2},Y_{7}\right]  =3Y_{7}$ & \\

& $\left[Y_{2},Y_{i}\right] =Y_{i+2}, i=3,5$&  & \\

$\frak{g}_{8}^{28}$ & $\left[  V_{1},Y_{i}\right]

=iY_{i}, i=1,2,3,4,5,7$ & $\left[V_{2},Y_{i}\right] =2Y_{i}, i=3,4$ & $d\omega_{y_{1}}+d\omega_{y_{7}}$\\

& $\left[V_{2},Y_{2}\right] =Y_{2}$& $\left[  V_{2},Y_{5}\right]  =3Y_{5}$ & \\

& $\left[V_{2},Y_{7}\right] =4Y_{7}$& $\left[  Y_{1},Y_{3}\right]  =Y_{4}$ & \\

& $\left[Y_{2},Y_{i}\right] =Y_{i+2}, i=3,5$& $\left[  Y_{3},Y_{4}\right]  =-Y_{7}$ & \\

$\frak{g}_{8}^{29}$ & $\left[  V_{1},Y_{i}\right]

=iY_{i}, i=1,2,3,4,5$ & $\left[V_{2},Y_{i}\right] =Y_{i}, i=2,3,4$ & $d\omega_{y_{4}}+d\omega_{y_{5}}$\\

& $\left[V_{1},Y^{\prime}_{3}\right] =3Y^{\prime}_{3}$& $\left[  V_{2},Y^{\prime}_{3}\right]  =Y^{\prime}_{3}$ & \\

& $\left[V_{2},Y_{5}\right] =2Y_{5}$& $\left[  Y_{1},Y_{2}\right]  =Y_{3}$ & \\

& $\left[Y_{1},Y^{\prime}_{3}\right] =Y_{4}, i=3,5$& $\left[  Y_{1},Y_{3}\right]  =Y_{4}$ & \\

& $\left[Y_{2},Y_{3}\right] =Y_{5}$&  & \\

$\frak{g}_{8}^{30}$ & $\left[  V_{1},Y_{i}\right]

=iY_{i}, i=1,2,3,4,5$ & $\left[V_{3},Y_{i}\right] =Y_{i}, i=2,3,4$ & $d\omega_{y_{3}}+d\omega_{y_{4}}+d\omega_{y_{5}}$\\

& $\left[V_{2},Y_{5}\right] =Y_{5}$& $\left[Y_{1},Y_{i}\right] =Y_{i+1}, i=2,3$ & \\

$\frak{g}_{8}^{31}$ & $\left[  V_{1},Y_{i}\right]

=iY_{i}, i=1,2,3,4,5$ & $\left[V_{2},Y_{i}\right] =Y_{i}, i=2,3$ & $d\omega_{y_{2}}+d\omega_{y_{3}}+d\omega_{y_{5}}$\\

& $\left[V_{3},Y_{i}\right] =Y_{i}; i=4,5$& $\left[Y_{1},Y_{i}\right] =Y_{i+1}, i=2,4$ & \\\hline

\end{tabular}
\end{table}


\newpage

\begin{thebibliography}{9}

\bibitem{So} J.-M. Souriau. Structure des syst\`emes dynamiques, Dunod, Paris, 1970.

\bibitem{Fo} A. T. Fomenko, V. V. Trofimov. Integrable systems on Lie algebras and symmetric spaces, Gordon and Breach, New York 1989.

\bibitem{Un} G. Unal. Algebraic integrability and generalized symmetries of dynamical systems, Phys. Lett. \textbf{A 260} (1999), 352-359. 

\bibitem{CG} R. Campoamor-Stursberg, F. G. Gasc\'on and D. Peralta. Dynamical systems embedded into Lie algebras, J. Math. Phys. \textbf{42} (2001), 5741-5752.

\bibitem{Chu} B.-Y. Chu. Symplectic homogeneous spaces, Trans. Amer. Math. Soc. \textbf{197} (1974), 145-159.

\bibitem{Ge} M. Gell-Mann  and Y. Ne'eman. The eightfold way, Benjamin, New York, 1964.

\bibitem{Ar} A. Arima and F. Iachello. Collective nuclear states as representations of an $SU(6)$ group, Phys. Rev. Lett. \textbf{35} (1975), 1069-1072.

\bibitem{PWZ} J. Patera, R. T. Sharp, P. Winternitz and H. Zassenhaus. Invariants of real low dimension Lie algebras, J. Math. Phys. \textbf{17} (1976), 986-994.

\bibitem{Ne} Y. Ne'eman and Dj. \v Sija\v ci. Hadrons in an $\overline{\rm SL}(4,R)$ classification, Phys. Rev. D \textbf{37} (1988), 3267-3283.

\bibitem{Ha} J. Hano. On K\"ahlerian homogeneous spaces of unimodular Lie groups, Amer. J. Math. \textbf{79} (1957), 885-900.

\bibitem{PN} J. Pecina-Cruz. An algorithm to calculate the invariants of any Lie algebra,  J. Math. Phys. \textbf{35} (1994), 3146-3162.  

\bibitem{Bas} P. Basarab-Horwath, V. Lahno and R. Zhdanov. The structure of Lie algebras and the classification problem for partial differential equations, Acta Math. Appl. \textbf{69} (2001), 43-94.

\bibitem{Car} J. F. Cari\~{n}ena, A. Ibort, G. Marmo and A. Perelomov. On the geometry of Lie algebras and Poisson tensors,  J. Phys. A: Math. Gen. \textbf{27} (1994), 7425-7449.

\bibitem{Mu1} G. M. Mubarakzyanov. Classification of solvable Lie algebras of sixth order with a non-nilpotent basis element, Izv. Vys\v s. U\v cebn. Zaved. Matematika \textbf{35} (1963), 104-116.

\bibitem{Mu2} G. M. Mubarakzyanov. On solvable Lie algebras, Izv. Vys\v s. U\v cehn. Zaved. Matematika \textbf{32} (1963), 114-123.

\bibitem{Tu} P. Turkowski. Solvable Lie algebras of dimension six. J. Math. Phys. \textbf{31} (1990), 1344-1350.

\bibitem{Nd} J. C. Ndogmo. Invariants of solvable Lie algebras of dimension six. J. Phys. A \textbf{33} (2000), 2273-2287.

\bibitem{Tu2} P. Turkowski. Low-dimensional real Lie algebras, J. Math. Phys. \textbf{29} (1988), 2139-2144. 

\bibitem{Tu3} P. Turkowski. Structure of real Lie algebras, Linear Algebra Appl. \textbf{171} (1992), 197-212.

\bibitem{AG} J. M. Ancochea Bermudez and M. Goze. On the classification of rigid Lie algebras, J. Algebra \textbf{245} (2001), 68-91.

\bibitem{Ca} R. Campoamor-Stursberg. Invariants of solvable rigid Lie algebras up to dimension $8$, J. Phys. A \textbf{35} (2002), 6293-6306.

\bibitem{Ca2} R. Campoamor-Stursberg. On the invariants of some solvable rigid Lie algebras, J. Math. Phys., to appear.

\bibitem{Ca3} R. Campoamor-Stursberg. Non-semisimple Lie algebras with Levi factor $\frak{so}(3),
\frak{sl}(2,\mathbb{R})$ and their invariants, J. Phys. A: Math. Gen., to appear.

\bibitem{G1} M. Goze. Mod\`eles d'alg\`ebres de Lie frobeniusiennes, C. R. A. S. Paris S\'er. I \textbf{293} (1981), 425-427.

\bibitem{Ahn} N. H. Ahn and L. Van Hop. Le centre de l'alg\`ebre enveloppante du produit semi-direct de l'alg\`ebre de Heisenberg et une alg\`ebre r\'eductive, C. R. A. S. Paris S\'er. I \textbf{303} (1986), 783-786.


\bibitem{Ca5} J. M. Ancochea, R. Campoamor-Stursberg. Symplectic forms and  products by generators, Comm. Algebra \textbf{30} (2002), 4235-4249.

\bibitem{AC1} J. M. Ancochea, R. Campoamor-Stursberg. On the cohomology of frobeniusian model Lie algebras, Forum Math., to appear.

\bibitem{BaN} H. Bacry, J. Nuyts. Classification of ten-dimensional kinematical groups with space isotropy, J. Math. Phys. \textbf{27} (1986), 2455-2457.

\bibitem{Sz} M. Szydlowski. The problem of chaotic behaviour in a homogeneous arbitrarily dimensional cosmology, Phys. Lett. \textbf{B 195} (1987), 31-35.
\end{thebibliography}
\end{document}